# Hippocluster: an efficient, hippocampus-inspired algorithm for graph clustering

Eric Chalmers, Artur Luczak

**Abstract**—Random walks can reveal communities or clusters in networks, because they are more likely to stay within a cluster than leave it. Thus, one family of community detection algorithms uses random walks to measure distance between pairs of nodes in various ways, and then applies K-Means or other generic clustering methods to these distances. Interestingly, information processing in the brain may suggest a simpler method of learning clusters directly from random walks. Drawing inspiration from the hippocampus, we describe a simple two-layer neural learning framework. Neurons in one layer are associated with graph nodes and simulate random walks. These simulations cause neurons in the second layer to become tuned to graph clusters through simple associative learning. We show that if these neuronal interactions are modelled a particular way, the system is essentially a variant of K-Means clustering applied directly in the walk-space, bypassing the usual step of computing node distances/similarities. The result is an efficient graph clustering method. Biological information processing systems are known for high efficiency and adaptability. In tests on benchmark graphs, our framework demonstrates this high data-efficiency, low memory use, low complexity, and real-time adaptation to graph changes, while still achieving clustering quality comparable to other algorithms.

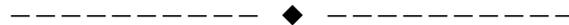

## 1 INTRODUCTION

The world contains many kinds of networks: physical road networks, the network of interconnected web pages in the worldwide web, social and professional networks, and ecological networks, to name a few. Networks are part of how the world is organized. A network can be represented and modelled using a *graph* - a data structure made up of objects called "nodes" or "vertices", connected by links called "edges". Wherever graphs are used it is often desirable to partition the graph into clusters or communities: groups of nodes more densely connected to each other than to other nodes. Analyzing the graph structure in this way is called *graph clustering* or *community detection*.

There is interesting and active research around graph clustering methods. For a good overview we refer the reader to the multidisciplinary review by Javed et al [1] or the recent review by Al-Andoli et al [2]; this paper will focus on methods based on random walks and/or K-Means clustering [3]. In general, generic clustering algorithms such as K-Means - used for clustering datapoints in a multidimensional space where concepts like *distance* and *mean* have clear meaning - cannot be directly applied to graphs. Sieranoja & Fränti point out, for example, that "a k-means based [graph clustering] solution is still missing... [because] k-means requires calculating the mean of a cluster, which is seemingly not possible for graphs." [4] Thus, one approach to graph clustering tries to *make* generic methods applicable. This has been attempted in at least three ways:

1) *Embedding the graph in a continuous space where generic clustering applies*. Embedding techniques try to arrange the graph's nodes and edges on a surface in a way that preserves most of the graph's structure. This surface then becomes a vector space where conventional clustering approaches apply. Embedding transformations can be effective in some applications [5] but Tandon et. al point out they are less practical than algorithms that cluster graph nodes directly [6].

2) *Adapting generic clustering methods for clustering graph nodes directly*. For K-Means, this can mean substituting some graph-centric metric in place of conventional (e.g. Euclidean) distance measurements. Previous work has redefined distance in terms of connections to a cluster's nodes [7], node degree centrality [8], shared nodes between overlapping clusters [9], or other graph metrics [10], [11]. In 2022 Sieranoja & Franti were perhaps the first to successfully adapt the core K-Means mechanism itself for graph clustering [4]. Experiments in this paper will include their algorithm as a representative of the family of K-Means based graph clustering methods.

3) *Explicitly computing similarity or distance-like scores between pairs of nodes and then clustering on those scores*. Of particular interest here is the use of random walks to measure similarity between nodes. A random walk through a graph will tend to stay inside clusters, since by definition there are more edges leading back into the cluster than out of it. Thus, random walks tend to reveal something about the graph structure; that is, nodes which frequently appear in the same random walks can be considered similar, and probably belong to the same com-

---

*Authors are with the Canadian Centre for Behavioural Neuroscience, University of Lethbridge, 4401 University Drive, Lethbridge, Alberta, T1K3M4, Canada*

- *Eric Chalmers E-mail: dchalmer@ualberta.ca*
- *Artur Luczak E-mail: luczak@uleth.ca*

munity. This property was articulated beautifully by Rosvall and Bergstrom [12], though it has been exploited in various ways by algorithms as early as PageRank [13]. van Dongen's MCL algorithm was an early application of the property for graph clustering, using a linear algebra approach to simulate all possible random walks simultaneously [14]. Later algorithms generated random walks through graphs in order to measure distance between nodes [15]–[18] or simulate propagation of each node's influence through the graph [19] for clustering purposes. Liu et al. found good results using random-walk-based distance measures for K-Means clustering specifically [20]. Recent work attempts to improve the random walks themselves, making them more adaptive [21], or more likely to respect cluster boundaries [22]–[24]. We include the Restrained Random-walk Similarity method of Okuda at al. in our experiments, as a recent and high-impact representative of the family of random-walk-based graph clustering methods.

One theme underlying past research is the addition of some extra machinery between the graph and the clustering method: i.e. using graph measurements or random walks to compute or measure distance/similarity between nodes, so that a clustering approach like K-Means can *then* be applied. In this paper we briefly review some phenomena observed in the Hippocampus - a brain structure involved in learning and memory - and propose that they suggest a more elegant solution: learning a clustering directly from random walks themselves.

### 1.1 Contributions of this paper

- We show that, while previous work has added extra computational machinery to adapt clustering methods like K-Means to graph clustering, it is possible to use Spherical K-Means to learn a clustering directly from random walks, and that this elegant approach grows naturally from a brain-inspired abstraction framework.

- We implement this brain-inspired framework and test it on benchmark graphs. We demonstrate its clustering efficacy, but also its high data, memory, and time efficiency. We also show that it can adapt to changes in the graph.

## 2 LEARNING FRAMEWORK

### 2.1 The neural inspiration for Hippocluster

The goal of graph clustering is to aggregate nearby nodes into communities or clusters that reflect the graph's inherent structure. These clusters can then serve as an abstract representation of the graph, in much the same way that K-means cluster centroids can become prototypes that serve as an abstract or simplified representation of a conventional dataset.

Similar hierarchical abstractions have been observed in the Hippocampus, which contains *place cells* that become tuned to represent specific places in an environment and fire whenever the animal is there. Place cells at the dorsal end of the hippocampus become tuned to small, specific places, while those at the ventral end become tuned to larger, more general areas [25] that could encompass many specific places (Fig. 1). Thus, the Hippocampus seems to encode spatial information (and probably information generally [26], [27]) using hierarchical abstraction. Our previous work showed that this hierarchical abstraction may be important for efficient and adaptable, human-like reinforcement learning [28], [29].

One interesting feature of these dorsal place cells is the *Replay* phenomenon - in which place cells fire to rapidly play out remembered or imagined trajectories. Whatever else these replays are doing, could they also help form hierarchical abstractions? Random walks through a graph contain information about the graph's structure [12], so these replayed trajectories through dorsal (specific) place cells would also contain information about the environment they traverse. They could therefore give rise in some way to abstract representations of places that respect environmental boundaries.

To illustrate this, we imagine a population of place cell neurons assigned to represent specific places, and a sec-ond layer of neurons that are each tuned to respond to a different (initially random) subset of these place cells. Simulated random trajectories activate the corresponding set of place cells, and through repeated simulation of such trajectories each higher-level neuron gradually becomes tuned to a contiguous set of specific places. This is because the string of places in the trajectory must be accessible from each other, while places distant from each other will not appear together in a given trajectory. With each simulated trajectory, the second-level neuron that responds the strongest has its connections to those place cells strengthened through simple associative learning ("cells that fire together wire together") and its connections to other places weakened. It is then even more likely to respond to that trajectory or an overlapping one next time, and less likely to respond to a trajectory covering some other region. Over time this simple associative learning process produces second-level neurons that aggregate place cells into groups respecting the environment's layout.

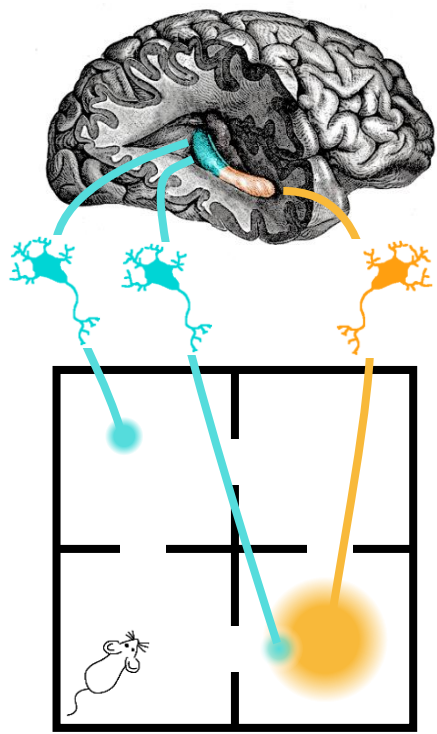

Fig. 1. Illustration of "place cells" in the Hippocampus. Cells at one end of the Hippocampus represent small regions (depicted in blue) while those at the other end represent larger ones (yellow). Similarly, our Hippocluster algorithm uses one set of neurons to represent specific graph nodes (small regions), and another representing clusters (large regions).

Now to connect this idea to the graph context. We can let the place cells represent individual graph nodes, the second-level cells represent clusters, and the connection weights between them indicate relative cluster membership. The learning process consists of simulating many random walks through the graph and adjusting the connection weights incrementally after each. This process gradually realizes the goal of aggregating nodes into appropriate clusters simply by observing random walks, with little other computational machinery or memory needed.

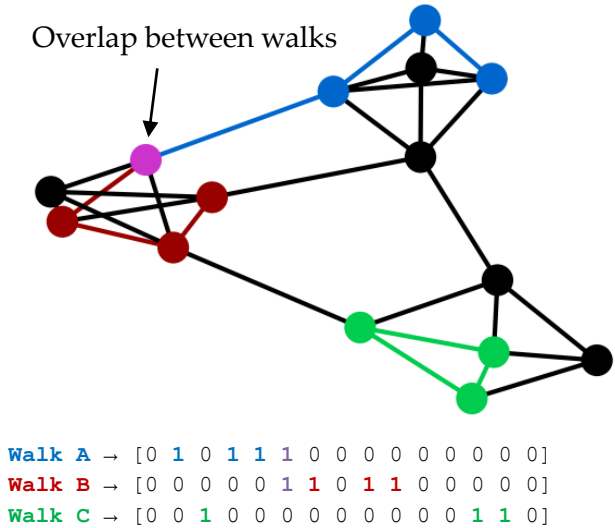

```
Walk A → [0 1 0 1 1 1 0 0 0 0 0 0 0 0 0]
Walk B → [0 0 0 0 0 1 1 0 1 1 0 0 0 0 0]
Walk C → [0 0 1 0 0 0 0 0 0 0 0 1 1 0]
```

Fig. 2. Representing random walks as length-N binary vectors. Hippocluster uses a bijective map to assign graph nodes to vector elements. A random walk vector uses this map to set elements indicating which nodes were visited. These vectors form an N-dimensional binary space in which all graph nodes are equidistant, each random walk is a point, and random walks only have measureable (cosine) similarity if they overlap. In this space, nodes no longer contain information about graph structure – all that information resides in random walks. In this example length-4 walks A and B have a nonzero cosine similarity because they overlap at one node, but both are equidistant (cosine similary of zero) from walk C.

## 2.2 Hippocluster as a special application of Online Spherical K-Means

The previous section imagined two layers of neurons, with neurons in the first layer assigned to represent specific places or graph nodes, while neurons in the second layer become tuned to represent clusters. A random walk (through the graph) activates the first-layer neurons along its route, creating a particular input pattern in the first layer. If the graph has N nodes, we can model this input pattern as a length-N, sparse binary vector with an element for each node, and the node elements visited by the random walk set to one. Encoding random walks in this way represents the graph as an N-dimensional space in which each dimension corresponds to one graph node (Fig. 2). For example, consider a simple graph with four nodes N={A,B,C,D}. We create a four-dimensional space in which the first dimension represents A, the second represents B, etc. Thus the vertex A is represented by the vector [1,0,0,0], D is represented as [0,0,0,1], etc.

There are two important things to note about this sparse high-dimensional space. First, any random walk, if seen as a set of its constituent nodes, can be represented as a single point in this space (the walk {A B}, {B C}, {C A} becomes the vector [1,1,1,0]). Trajectories with a high degree of overlap will then have high cosine similarity. Second, all nodes are equidistant in this space: the vector representations for any two nodes have a cosine similarity of zero (or a Euclidean distance of 1.41) between them, regardless of how close or distant they were in the original graph. This means that all information about graph structure now resides in the random walk vectors.

The learning process consists of identifying the "winning" second-layer neuron that reacts most strongly to each input pattern, and incrementally strengthening its connection to that pattern (see Fig. 3). We model the relationship between second-layer neurons' activity and first-layer neurons' activity as a dot product between the input pattern and the connection weights:

$$\mathbf{y} = \mathbf{W} \cdot \mathbf{x}^T \quad (1)$$

Where $\mathbf{y}$ is the vector of second-layer neuron activations, $\mathbf{W}$ is a sparse matrix of connection weights, with $\mathbf{W}_{i,j}$ being the weight of the connection between the $j^{th}$ first-layer neuron and $i^{th}$ second-layer neuron, and $\mathbf{x}$ is the vector of first-layer neuron activations (the input pattern).

The second-layer neuron with the largest response to a given input pattern becomes more closely associated with that pattern, in winner-takes-all fashion. We model this learning process as a push of that neuron's weights toward the input pattern. If the index of the winning neuron is denoted $c$:

$$c = \mathrm{argmax}(\mathbf{y}) \quad (2)$$
$$\mathbf{W}_c = (1-\eta)\mathbf{W}_c + \eta\mathbf{x} \quad (3)$$

where $\mathbf{W}_c$ is the $c^{th}$ row of $\mathbf{W}$, containing the weights of connections to the winning ($c^{th}$) second-layer neuron. $\eta$ is a learning rate parameter.

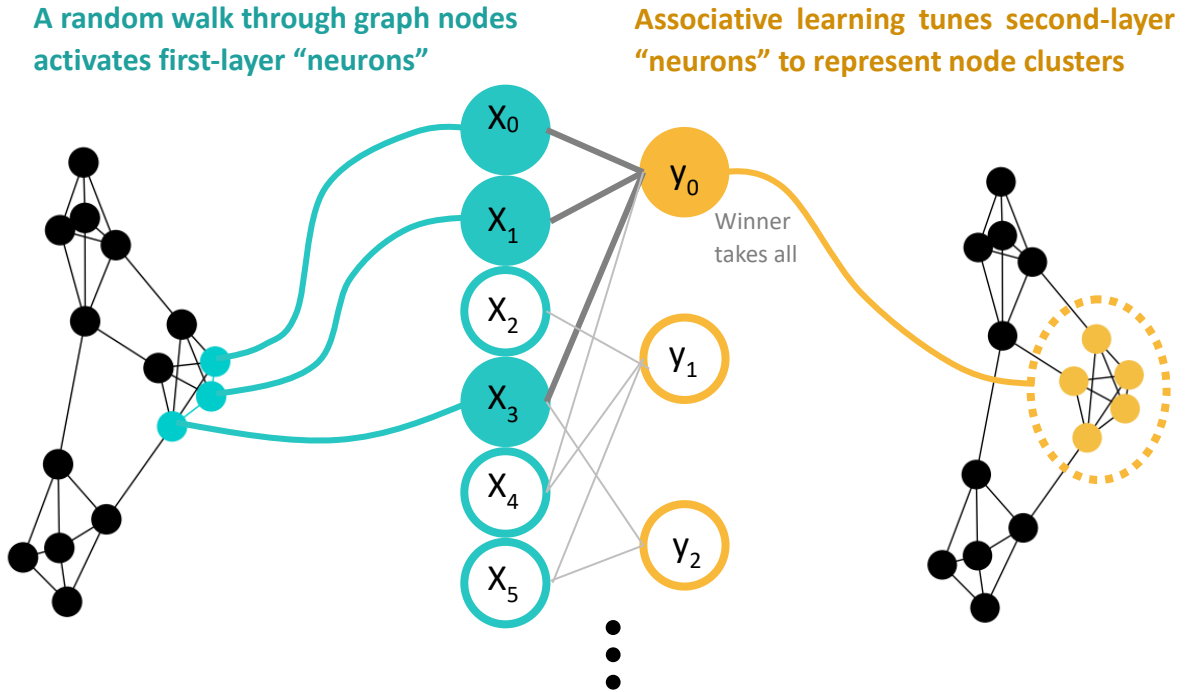

Fig. 3. Hippocluster is a 2-layer learning system. Neurons in the first layer each represent a graph node. As random walks through the graph activate these neurons, they in turn activate a second-layer neuron in winner-takes-all fashion, and this winning neuron becomes more closely tuned to that walk. Seond-layer neurons gradually become tuned to graph clusters, in which many walks tend to overlap (because walks only have measureable similarity if they overlap).

At this point our framework can be nearly identified with Online Spherical K-Means clustering [30]. Spherical K-Means Clustering is a variant of K-Means clustering that has been widely applied in document clustering, where documents are represented by term frequency vectors that indicate occurrences of particular words or phrases in a document [31], just as our trajectory vectors indicate the occurrence of particular nodes in a random walk. In both cases the vectors are sparse, and since the Euclidean distance calculation used in conventional K-Means can be sensitive to sparsity differences, Spherical K-Means uses Cosine similarity instead. For efficiency, it also projects cluster centroids and data points onto a unit hypersphere, where the Cosine similarity calculation becomes a simple dot product. *Online* Spherical K-Means [30] is a sequential adaptation of Spherical K-Means that operates on one sample at a time, by nudging the closest-match centroid toward the new sample according to some learning rate. So far then, our framework is simply Online Spherical K-Means clustering applied in the walk-space; with random-walk vectors used as inputs, and the connection weights between first- and second-layer neurons being the cluster centroids.

### 2.3 Adapting dynamically to graph changes

Neurons in the brain form new connections and lose old ones continuously, in a lifelong learning process. Similarly, we can allow the number of neurons and connections in our framework to change dynamically. This lets us handle graph changes dynamically in real-time.

We extend the Online Spherical K-Means approach with a bijective map, which represents and tracks the assignments of first-layer neurons to particular graph nodes. Just as Hippocampus place cells are recruited to represent new places, this bijective map expands to include new graph nodes as they are encountered. The weight matrix **W** expands to match the map's length, allowing for new connections to form between the new first-layer neurons and the existing second-layer ones (i.e. allowing the new nodes to be annexed by existing clusters).

If a graph node stops appearing in random walks, the process of updating and reprojection will eventually drive the corresponding weights (i.e. the corresponding column of **W**) close to zero. We further extend the algorithm by setting to zero and pruning any weights that fall below a minimum threshold - representing the degeneration of unused connections over time. This allows lost nodes to be "forgotten" so the clustering may adapt dynamically to graph changes, and has the additional benefit of keeping the connections in the weight matrix sparse for efficiency.

Finally, we take the engineering liberty of processing inputs in small batches (rather than strictly one-at-a-time) as recommended by Sculley [32]. The full algorithm is shown in Algorithm 1.

Algorithm 1. The Hippocluster learning framework

| | |
|---|---|
| Input: | graph $G$ (containing $N$ nodes)<br>batch size $b$<br>random walk length $l$<br>total budget for random walks $max\_walks$<br>learning rate $\eta$<br>connection weight threshold $t$ |
| Output: | mapping of graph nodes to cluster assignments |
| Initialize: | bijective map $map$: node $\mapsto$ integer<br>$K \times 0$ sparse connection weight matrix $\mathbf{W}$<br>($\mathbf{W}_{i,j}$ is weight of connection between the $j^{th}$ first-layer neuron and $i^{th}$ second-layer neuron) |

Steps:

1. Sample a size-$b$ batch of length-$l$ random walks from $G$: {$walk_0, walk_1 \ldots walk_b$}

2. add newly encountered nodes to the learning system.
   for each new node:
   $map$[node] = 1 + size of $map$
   expand $\mathbf{W}$ to have width = size of map

3. convert each $walk_i$ to a sparse unit vector $\mathbf{x}_i$
   (i.e. $\mathbf{x}_{i,j}$ = 1 / sqrt( |$walk_i$| ) where $j$ = $map$[node] for each node in $walk$. This makes |$\mathbf{x}_i$| = 1)

4. if first iteration, select $k$ walk vectors to form the rows of $\mathbf{W}$

5. for each random walk vector, identify index $c$ of the most strongly-activated second-layer neuron, per the connection weights in $\mathbf{W}$
   for i = 0:$b$
   $c_i$ = argmax($\mathbf{W} \bullet \mathbf{x}_i^T$)

6. tune each second layer neuron to more closely match the input patterns that activated it.
   for i = 0:$b$
   $\mathbf{W}_{c_i}$ = (1 - $\eta$)$\mathbf{W}_{c_i}$ + $\eta \mathbf{x}_i$

7. reproject connection weights onto unit sphere by L2-normalizing rows of $\mathbf{W}$

8. remove weak connections by setting values in $\mathbf{W}$ less than $t$ to zero

   repeat steps 1-8, $max\_walks/b$ times

9. map each node in $G$ to a cluster, (i.e. determine which second-layer neuron each first-layer neuron is connected to most strongly)

   for node in $G$:
   $n$ = $map$[node]
   node $\mapsto$ argmax$_c$ ($\mathbf{W}_{c,n}$)

## 3 EXPERIMENTS

The Hippocluster algorithm was tested on Lancichinetti–Fortunato–Radicchi (LFR) benchmark graphs [33] of varying sizes, using LFR parameters of $\gamma$=2, $\beta$=1.1, $\mu$=0.1, $s_{min}$=20, $s_{max}$=50 (similar settings to those used by Okuda et. al [24]). LFR graphs are widely-used benchmarks as they closely resemble real-world networks, and so allow controlled but realistic experiments. Hippocluster's performance was compared to that of two recently-proposed comparators: the Restrained Random-walk Similarity method (RRWS) of Okuda et al. [24] (a recent random-walk-based method) and the K-Algorithm of Sieranoja & Fränti [4] (a recent application of K-means to graph clustering).

RRWS executes multiple random walks from each node and records a set of nodes which occur frequently across these walks. The Jaccard similarity between the sets for each node-pair is measured, and a threshold applied to these similarities to create a community graph. It is a good representative of the many algorithms that use random walks to estimate distance/similarity between nodes, and then derive clusters based on these estimates - but it improves on these methods by constraining the walks in a clever way, making them less likely to leave a cluster.

The K-Algorithm is a sequential adaptation of K-means intended for graph clustering. It iteratively shuffles nodes between clusters, accepting changes which improve a novel quality metric. We note the original paper contains important new methods for initializing the clusters, and for improving the clustering result. We omit these extensions from our experiments, as they are generally applicable and could be used to improve many clustering algorithms (including ours). That is, for our experiments we test the core clustering algorithm, and try to control for initialization and post-processing. In our experiments the K-Algorithm is initialized using the same K-Means++ method [34] as Hippocluster.

### 3.1 Cluster quality

Parameters for each algorithm were tuned in a grid-search on a "training" LFR graph, and the best parameters applied on a separate "test" LFR graph created using the same graphs parameters but a different random seed. The lengths of Hippocluster's random walks were randomly-selected between 20 and 75 (covering the range of cluster sizes, and extending 50% beyond the maximum size). Each discovered clustering was evaluated in terms of Normalized Mutual Information score (NMI). NMI measures the amount of information shared between two variables, and in this case expresses the similarity between the discovered clustering and the ground-truth LFR communities as a score between 0 (no correspondence between the groupings) and 1 (perfect correspondence). Given a set of ground-truth community assignments $Y$ and a corresponding set of cluster assignments $\hat{Y}$, mutual information (MI) is defined as:

[35]
$$\mathrm{MI}(Y; \hat{Y}) = \sum_{y \in Y} \sum_{\hat{y} \in \hat{Y}} P_{(Y,\hat{Y})}(y, \hat{y}) \cdot \log\left(\frac{P_{(Y,\hat{Y})}(y, \hat{y})}{P_Y(y) \cdot P_{\hat{Y}}(\hat{y})}\right) \quad (4)$$

where $P_{(Y,\hat{Y})}$ is the joint probability mass function of $Y$ and $\hat{Y}$, and $P_Y(y)$ and $P_{\hat{Y}}(\hat{y})$ are the marginal probabilities. For this paper, normalized mutual information (NMI) normalizes MI by the arithmetic mean of the entropies $H(Y)$ and $H(\hat{Y})$:

$$\mathrm{NMI}(Y; \hat{Y}) = \frac{2 \cdot \mathrm{MI}(Y; \hat{Y})}{H(Y) + H(\hat{Y})} \quad (5)$$

Fig. 4 shows the algorithms' clustering performance in terms of NMI.

### 3.2 Memory efficiency

To test memory efficiency, we measured the final memory requirements of each algorithm. We assumed the use of sparse matrices and so measured the maximum number of non-zero float values stored in Hippocluster's weight/centroid matrix, in RRWS's community graph, and in K-Algorithm's cluster assignments. Note that the size of the graph itself or the memory required to store the graph's adjacency matrix is excluded. Storing only the current cluster assignments makes K-Algorithm highly memory efficient, as shown in Fig. 5.

### 3.3 Execution time

Running time was measured for LFR benchmark graphs of various sizes. For these tests the RRWS algorithm was allowed 38×N random walks, and the Hippocluster algorithm was allowed 4×N (the values that allowed each algorithm to achieve ~95% of their maximum NMI in the Data Efficiency tests). Results are plotted in Fig. 6.

All algorithms were implemented in Python, and we believe the implementations of RRWS and K-Algorithm are faithful to the presentations in the original papers. However, it is possible that any (or all three) of our implementations could be improved, potentially changing these results. Furthermore, we note that the RRWS algorithm could be accelerated by processing nodes in parallel, and Hippocluster could be similarly accelerated through a parallel implementation of K-means as in [35]. Thus, it should be acknowledged that these results apply only to our particular implementations.

### 3.4 Data efficiency

We measure data efficiency for Hippocluster and RRWS by measuring the number of random walks needed to reach a given percentage of the best-achieved NMI. As shown in Fig. 7, Hippocluster converges to a good clustering solution in relatively few random walks. RRWS requires more walks to be generated to reach the same level of performance.

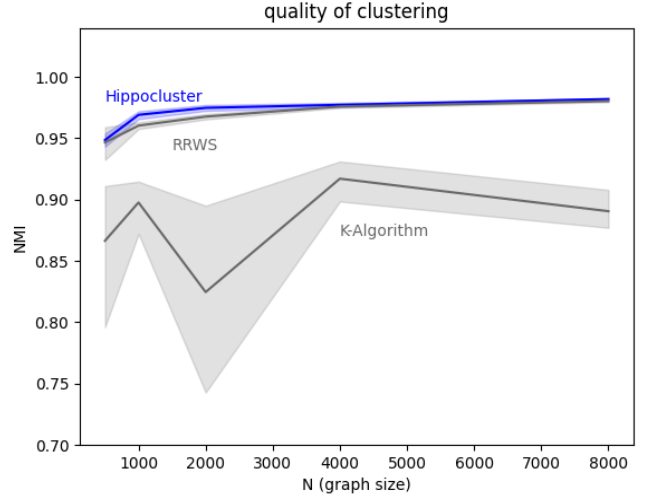

Fig. 4. NMI over 10 runs on LFR graphs of various sizes. Shaded regions show 95% confidence intervals. Hippocluster performs as well or better than RRWS and K-Algorithm

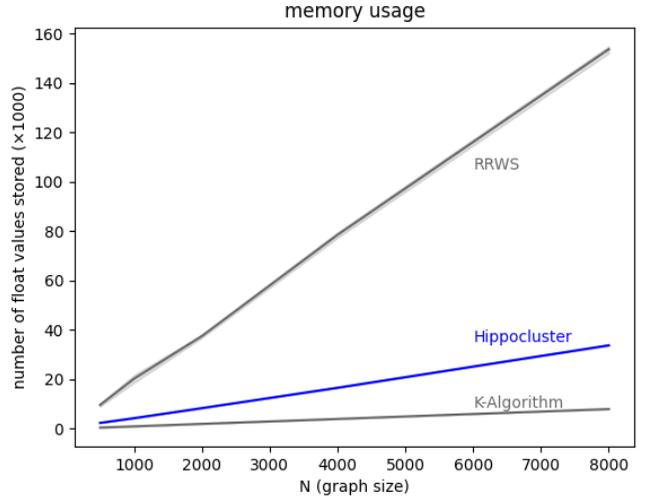

Fig. 5. Memory usage across LFR graphs of various size. K-Algorithm requires the least memory, but Hippocluster's memory requirements are also quite low.

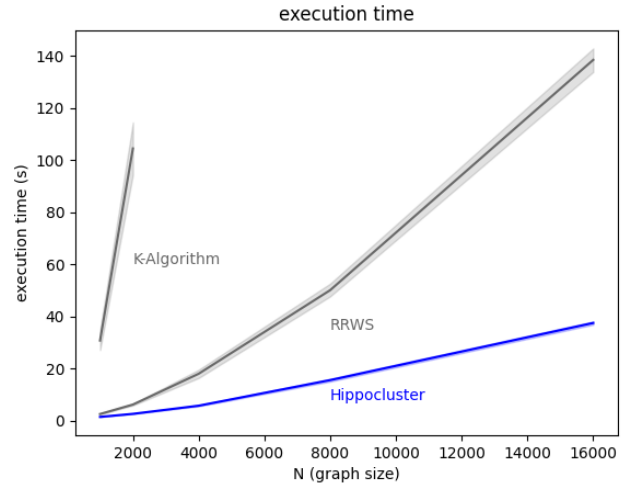

Fig. 6. Execution time for graphs of various sizes.

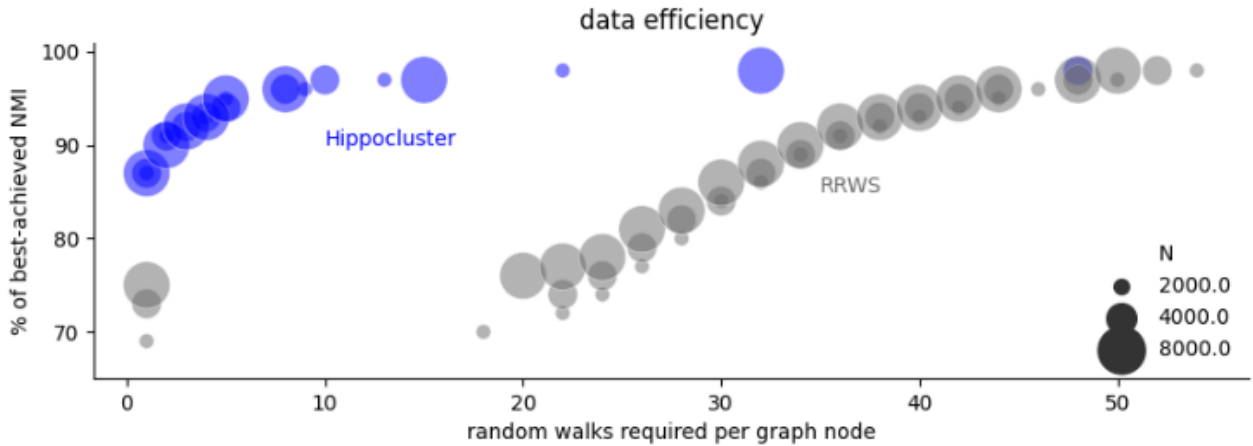

Fig. 7. NMI after seeing a given number of random walks. Circle size indicates graph size (*N*). Hippocluster acheives 95% of its best NMI after processing about 4 walks per node, while RRWS needs about 38, making Hippocluster more data efficient.

### 3.5 Adapting to graph changes

To demonstrate Hippocluster's ability to handle graph changes, we modify an LFR graph during Hippocluster's learning process. Graphs of 2000, 3000, and 4000 nodes were generated. After updating Hippocluster with 100 batches of random walks, 10% of the graph's nodes were removed. After 100 more batches, the removed nodes were replaced with an equal number of new nodes, with new edges. The removals/additions were distributed roughly evenly across the graph's communities. We measured the NMI between the ground-truth and the current clustering through the process, with the results shown in Fig. 8.

## 4 DISCUSSION

Here we developed a novel and elegant method of graph clustering, by applying spherical K-Means directly to random walks. This approach falls naturally out of a biologically-inspired cluster-learning framework, which notes that abstraction or generalization of information (as in clustering) and simulation of trajectories are apparently both important mechanisms in the Hippocampus. Biologically-inspired models such as this can provide insight into brain function, and also improve the state-of-the-art in machine learning. This interplay (and the need for it) ha been described by Hassabis et al. [36], and we have benefited from it in our past explorations of hierarchical abstraction in reinforcement learning, [28], [29], transfer learning through egocentric and allocentric learning system cooperation [37], and how neurons may optimize energy balance by predicting their own future activity [38].

The framework imagines a simple two-layer neural learning system that gradually associates intra-cluster nodes together, as it is repeatedly exposed to random walks. By modelling this system's behavior in a particular way, we arrive at something very similar to online spherical K-Means clustering. Extending the framework with some dynamic capabilities allows it to adapt to graph changes.

This biologically-inspired approach compares very favorably against the Restrained Random-Walk Similarity (RRWS) method [24] and the K-Algorithm [4]. Our results suggest Hippocluster can achieve better performance than K-Algorithm, with significantly better memory and data efficiency than RRWS, and possibly faster execution than either. Hippocluster's fast execution comes from its data efficiency: comparing Fig. 6 and Fig. 7, we see that RRWS uses less time per random walk, but Hippocluster uses less time overall because it can find a good solution with relatively few random walks.

Hippocluster's high data efficiency is especially important in the biological context - where animals must be able to learn as quickly as possible - but also in the engineering context, where generating random walks through a very large graph could be computationally expensive. Hippocluster's memory efficiency comes from the fact that it need only maintain the sparse weight matrix. In the worst (dense) case, this matrix has size $O(Nk)$. In the best case it is sparse and binary with each first-layer neuron connected to exactly one second-layer neuron, and has size $O(N)$. K-Algorithm consumes the least memory, because it only stores the current cluster assignments for each node, and is therefore always $O(N)$. A similarity matrix like that constructed by RRWS can potentially store a value for every node-pari: it has size $O(N)$ in the best case of completely isolated clusters, but is up to $O(N^2)$ for a dense graph.

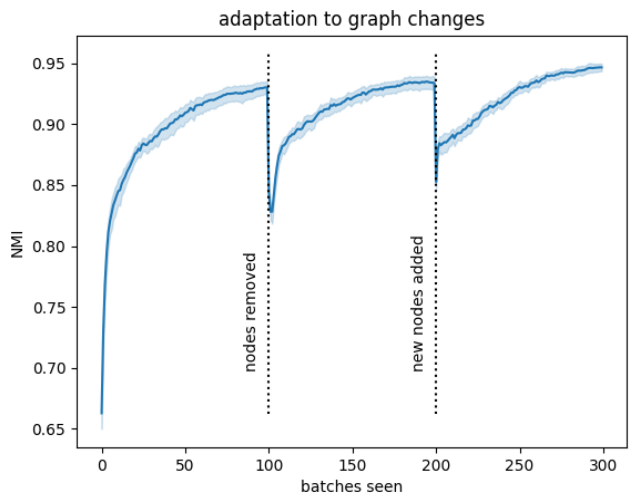

Fig. 8. NMI measured over time (i.e. random walk batches processed). After 100 batches, 10% of the graph nodes are removed After 100 more batches, those nodes are replaced with new nodes. The algorithm adapts to both changes.

K-Means clustering allows the user to select the number of clusters to discover, and Hippocluster inherits this feature. In practice there is often no "correct" number of clusters, as networks could be analyzed at various levels of granularity and communities may overlap. However, we can find a value of $k$ that best matches the inherent structure of the graph by executing the algorithm with various values of $k$, and choosing the value that optimizes some internal measure of cluster quality, such as Modularity [39].

There is an interesting comparison to be made between graph-embedding methods and our framework. Graph embedding attempts to position graph nodes in a continuous space in a way that preserves the graph's structure. Once in this space, the nodes can be clustered using any generic clustering algorithm. But the complexity of this embedding operation is usually high, making graph embedding methods somewhat impractical [6] and, in our experience, slow. In contrast, Hippocluster operates in the random-walk space - a binary space, or hypercube, where each dimension represents a graph node. Rather than preserving graph structure, this space actually erases it completely: making every node equidistant from every other node (in this space, the binary vector representations of any two nodes always have a cosine similarity of zero). All structural information now resides in the random walks, which appear at hypercube corners involving multiple dimensions (nodes). These random walks become the (only) thing that ties similar nodes together. Interestingly, forgoing the extra structure-preserving effort of graph embedding allows more-efficient but still highly effective clustering.

## CODE REPOSITORY

Ready-to-use Python code for Hippocluster can be found at https://github.com/echalmers/hippocluster

## REFERENCES

[1] M. A. Javed, M. S. Younis, S. Latif, J. Qadir, and A. Baig, "Community detection in networks: A multidisciplinary review," *J. Netw. Comput. Appl.*, vol. 108, pp. 87–111, Apr. 2018, doi: 10.1016/j.jnca.2018.02.011.

[2] M. N. Al-Andoli, S. C. Tan, W. P. Cheah, and S. Y. Tan, "A Review on Community Detection in Large Complex Networks from Conventional to Deep Learning Methods: A Call for the Use of Parallel Meta-Heuristic Algorithms," *IEEE Access*, vol. 9, pp. 96501–96527, 2021, doi: 10.1109/ACCESS.2021.3095335.

[3] J. MacQueen, "Some Methods for classification and Analysis of Multivariate Observations," *Proc. 5th Berkeley Symp. Math. Stat. Probab.*, vol. 1, pp. 281–297, 1967.

[4] S. Sieranoja and P. Fränti, "Adapting k-means for graph clustering," *Knowl. Inf. Syst.*, vol. 64, no. 1, pp. 115–142, Jan. 2022, doi: 10.1007/s10115-021-01623-y.

[5] M. Xu, "Understanding Graph Embedding Methods and Their Applications," *SIAM Rev.*, vol. 63, no. 4, pp. 825–853, Jan. 2021, doi: 10.1137/20M1386062.

[6] A. Tandon, A. Albeshri, V. Thayananthan, W. Alhalabi, F. Radicchi, and S. Fortunato, "Community detection in networks using graph embeddings," *Phys. Rev. E*, vol. 103, no. 2, p. 022316, Feb. 2021, doi: 10.1103/PhysRevE.103.022316.

[7] A. Bóta, M. Krész, and B. Zaválnij, "Adaptations of the k-Means Algorithm to Community Detection in Parallel Environments," in *2015 17th International Symposium on Symbolic and Numeric Algorithms for Scientific Computing (SYNASC)*, Sep. 2015, pp. 299–302. doi: 10.1109/SYNASC.2015.54.

[8] B. Cai, L. Zeng, Y. Wang, H. Li, and Y. Hu, "Community Detection Method Based on Node Density, Degree Centrality, and K-Means Clustering in Complex Network," *Entropy*, vol. 21, no. 12, Art. no. 12, Dec. 2019, doi: 10.3390/e21121145.

[9] Z. Zhou, Z. Xiao, and W. Deng, "Improved community structure discovery algorithm based on combined clique percolation method and K-means algorithm," *Peer--Peer Netw. Appl.*, vol. 13, no. 6, pp. 2224–2233, Nov. 2020, doi: 10.1007/s12083-020-00902-9.

[10] T. Van Laarhoven and E. Marchiori, "Local network community detection with continuous optimization of conductance and weighted kernel K-means," *J. Mach. Learn. Res.*, vol. 17, no. 1, pp. 5148–5175, Jan. 2016.

[11] W. Wang and W. N. Street, "A novel algorithm for community detection and influence ranking in social networks," in *2014 IEEE/ACM International Conference on Advances in Social Networks Analysis and Mining (ASONAM 2014)*, Aug. 2014, pp. 555–560. doi: 10.1109/ASONAM.2014.6921641.

[12] M. Rosvall and C. T. Bergstrom, "Maps of random walks on complex networks reveal community structure," *Proc. Natl. Acad. Sci.*, vol. 105, no. 4, pp. 1118–1123, Jan. 2008, doi: 10.1073/pnas.0706851105.

[13] S. Brin and L. Page, "The anatomy of a large-scale hypertextual Web search engine," *Comput. Netw. ISDN Syst.*, vol. 30, no. 1, pp. 107–117, Apr. 1998, doi: 10.1016/S0169-7552(98)00110-X.

[14] Dongen, S.M. van and University Utrecht, "Graph clustering by flow simulation." Accessed: Mar. 21, 2022. [Online]. Available: https://dspace.library.uu.nl/handle/1874/848

[15] W. Wang, D. Liu, X. Liu, and L. Pan, "Fuzzy overlapping community detection based on local random walk and multi-dimensional scaling," *Phys. Stat. Mech. Its Appl.*, vol. 392, no. 24, pp. 6578–6586, Dec. 2013, doi: 10.1016/j.physa.2013.08.028.

[16] H. Zhou, "Distance, dissimilarity index, and network community structure," *Phys. Rev. E*, vol. 67, no. 6, p. 061901, Jun. 2003, doi: 10.1103/PhysRevE.67.061901.

[17] H. Zhou and R. Lipowsky, "Network Brownian Motion: A New Method to Measure Vertex-Vertex Proximity and to Identify Communities and Subcommunities," in *Computational Science - ICCS 2004*, Berlin, Heidelberg, 2004, pp. 1062–1069.

doi: 10.1007/978-3-540-24688-6_137.

[18] P. Pons and M. Latapy, "Computing Communities in Large Networks Using Random Walks," in *Computer and Information Sciences - ISCIS 2005*, Berlin, Heidelberg, 2005, pp. 284–293. doi: 10.1007/11569596_31.

[19] Y. Yan, Y. Bian, D. Luo, D. Lee, and X. Zhang, "Constrained Local Graph Clustering by Colored Random Walk," in *The World Wide Web Conference*, New York, NY, USA, May 2019, pp. 2137–2146. doi: 10.1145/3308558.3313719.

[20] J. Liu, "Comparative Analysis for k-Means Algorithms in Network Community Detection," in *Advances in Computation and Intelligence*, Berlin, Heidelberg, 2010, pp. 158–169. doi: 10.1007/978-3-642-16493-4_17.

[21] Y. Xin, Z.-Q. Xie, and J. Yang, "An adaptive random walk sampling method on dynamic community detection," *Expert Syst. Appl. Int. J.*, vol. 58, no. C, pp. 10–19, Oct. 2016, doi: 10.1016/j.eswa.2016.03.033.

[22] K. Kitaura, R. Matsuo, and H. Ohsaki, "Random Walk on a Graph with Vicinity Avoidance," Jan. 2022, pp. 232–237. doi: 10.1109/ICOIN53446.2022.9687140.

[23] Y. Yi, L. Jin, H. Yu, H. Luo, and F. Cheng, "Density Sensitive Random Walk for Local Community Detection," *IEEE Access*, vol. 9, pp. 27773–27782, 2021, doi: 10.1109/ACCESS.2021.3058908.

[24] M. Okuda, S. Satoh, Y. Sato, and Y. Kidawara, "Community Detection Using Restrained Random-Walk Similarity," *IEEE Trans. Pattern Anal. Mach. Intell.*, vol. 43, no. 1, pp. 89–103, 2019, doi: 10.1109/TPAMI.2019.2926033.

[25] B. A. Strange, M. P. Witter, E. S. Lein, and E. I. Moser, "Functional organization of the hippocampal longitudinal axis," *Nat. Rev. Neurosci.*, vol. 15, no. 10, Art. no. 10, Oct. 2014, doi: 10.1038/nrn3785.

[26] D. Aronov, R. Nevers, and D. W. Tank, "Mapping of a non-spatial dimension by the hippocampal–entorhinal circuit," *Nature*, vol. 543, no. 7647, Art. no. 7647, Mar. 2017, doi: 10.1038/nature21692.

[27] R. M. Mok and B. C. Love, "A non-spatial account of place and grid cells based on clustering models of concept learning," *Nat. Commun.*, vol. 10, no. 1, Art. no. 1, Dec. 2019, doi: 10.1038/s41467-019-13760-8.

[28] E. Chalmers, A. Luczak, and A. J. Gruber, "Computational Properties of the Hippocampus Increase the Efficiency of Goal-Directed Foraging through Hierarchical Reinforcement Learning," *Front. Comput. Neurosci.*, vol. 10, 2016, Accessed: Mar. 21, 2022. [Online]. Available: https://www.frontiersin.org/article/10.3389/fncom.2016.00128

[29] E. Chalmers, E. B. Contreras, B. Robertson, A. Luczak, and A. Gruber, "Context-switching and adaptation: Brain-inspired mechanisms for handling environmental changes," in *2016 International Joint Conference on Neural Networks (IJCNN)*, Jul. 2016, pp. 3522–3529. doi: 10.1109/IJCNN.2016.7727651.

[30] S. Zhong, "Efficient online spherical k-means clustering," in *Proceedings. 2005 IEEE International Joint Conference on Neural Networks, 2005.*, Jul. 2005, vol. 5, pp. 3180–3185 vol. 5. doi: 10.1109/IJCNN.2005.1556436.

[31] K. Hornik, I. Feinerer, M. Kober, and C. Buchta, "Spherical k-Means Clustering," *J. Stat. Softw.*, vol. 50, pp. 1–22, Sep. 2012, doi: 10.18637/jss.v050.i10.

[32] D. Sculley, "Web-scale k-means clustering," in *Proceedings of the 19th international conference on World wide web*, New York, NY, USA, Apr. 2010, pp. 1177–1178. doi: 10.1145/1772690.1772862.

[33] A. Lancichinetti, S. Fortunato, and F. Radicchi, "Benchmark graphs for testing community detection algorithms," *Phys. Rev. E*, vol. 78, no. 4, p. 046110, Oct. 2008, doi: 10.1103/PhysRevE.78.046110.

[34] D. Arthur and S. Vassilvitskii, "k-means++: the advantages of careful seeding," in *Proceedings of the eighteenth annual ACM-SIAM symposium on Discrete algorithms*, USA, Jan. 2007, pp. 1027–1035.

[35] F. Othman, R. Abdullah, N. A. Rashid, and R. A. Salam, "Parallel K-Means Clustering Algorithm on DNA Dataset," in *Parallel and Distributed Computing: Applications and Technologies*, Berlin, Heidelberg, 2005, pp. 248–251. doi: 10.1007/978-3-540-30501-9_54.

[36] D. Hassabis, D. Kumaran, C. Summerfield, and M. Botvinick, "Neuroscience-Inspired Artificial Intelligence," *Neuron*, vol. 95, no. 2, pp. 245–258, Jul. 2017, doi: 10.1016/j.neuron.2017.06.011.

[37] E. Chalmers, E. B. Contreras, B. Robertson, A. Luczak, and A. Gruber, "Learning to Predict Consequences as a Method of Knowledge Transfer in Reinforcement Learning," *IEEE Trans. Neural Netw. Learn. Syst.*, vol. 29, no. 6, pp. 2259–2270, Jun. 2018, doi: 10.1109/TNNLS.2017.2690910.

[38] A. Luczak, B. L. McNaughton, and Y. Kubo, "Neurons learn by predicting future activity," *Nat. Mach. Intell.*, vol. 4, no. 1, Art. no. 1, Jan. 2022, doi: 10.1038/s42256-021-00430-y.

[39] M. E. J. Newman, "Modularity and community structure in networks," *Proc. Natl. Acad. Sci.*, vol. 103, no. 23, pp. 8577–8582, Jun. 2006, doi: 10.1073/pnas.0601602103.

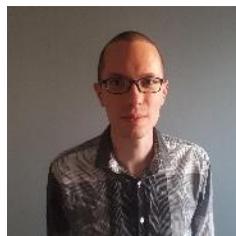

**Eric Chalmers** earned a BSc in electrical engineering in 2011, and a PhD in Electrical Computer Engineering in 2015, both from the University of Alberta. Eric's PhD was supported by grants from Alberta Innovates and the Women's and Children's Health Research Institute. He served as a post-doctoral fellow at the Canadian Center for Behavioral Neuroscience at the University of Lethbridge with support from the Natural Sciences and Engineering Research Council of Canada, before moving to industry and holding various data science and management positions. He now engages in a mix of academic and industry work.

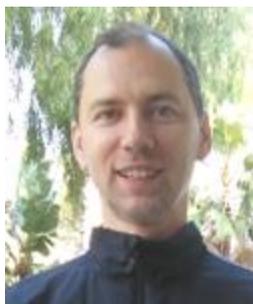

**Artur Luczak** received the M.Sc. degree in biomedical engineering from the Wroclaw University of Technology, Poland, and the Ph.D. degree from the Jagiellonian University, Krakow, Poland, which was complemented by the Marie Curie Fellowship at SISSA in Trieste, Italy. After a Post-Doctoral Fellowships at Yale and Rutgers University, he is currently a Professor at the Canadian Center for Behavioral Neuroscience, University of Lethbridge, AB, Canada. His laboratory is studying neural information processing using experimental and theoretical methods and how it is affected in different neurological disorders such as epilepsy. Dr. Luczak was elected to the College of the Royal Society of Canada in 2016.